# Hunting imaging biomarkers in pulmonary fibrosis: Benchmarks of the AIIB23 challenge

Yang Nan[a,b], Xiaodan Xing[a], Shiyi Wang[c], Zeyu Tang[a], Federico N Felder[b,c], Sheng Zhang[c], Roberta Eufrasia Ledda[d], Xiaoliu Ding[e], Ruiqi Yu[e], Weiping Liu[e], Feng Shi[f], Tianyang Sun[f], Zehong Cao[f], Minghui Zhang[g], Yun Gu[g], Hanxiao Zhang[g], Jian Gao[h], Pingyu Wang[i], Wen Tang[j], Pengxin Yu[j], Han Kang[j], Junqiang Chen[k], Xing Lu[l], Boyu Zhang[l], Michail Mamalakis[m], Francesco Prinzi[n], Gianluca Carlini[o], Lisa Cuneo[p], Abhirup Banerjee[q], Zhaohu Xing[r], Lei Zhu[r], Zacharia Mesbah[s,t], Dhruv Jain[s], Tsiry Mayet[s], Hongyu Yuan[u], Qing Lyu[u], Abdul Qayyum[c], Moona Mazher[v], Athol Wells[b,c], Simon LF Walsh[b,c,*], Guang Yang[a,b,c,w,*]

[a] Bioengineering Department and Imperial-X, Imperial College London, London, UK.
[b] Royal Brompton Hospital, London, UK.
[c] National Heart and Lung Institute, Imperial College London, London, UK.
[d] Department of Medicine and Surgery, University of Parma, Parma, Italy.
[e] Shanghai MicroPort MedBot (Group) Co., Ltd., China.
[f] Shanghai United Imaging Intelligence Co., Ltd., China.
[g] Institute of Medical Robotics, Shanghai Jiao Tong University, China.
[h] Department Computational Biology, School of Life Sciences, Fudan University, Shanghai, China.
[i] Cambridge International Exam Centre in Shanghai Experimental School, China
[j] InferVision Medical Technology Co., Ltd., China.
[k] Shanghai MediWorks Precision Instruments Co., Ltd, China.
[l] Sanmed Biotech Ltd., Zhuhai, China.
[m] University of Cambridge, Department of Psychiatry, Cambridge, UK.
[n] Department of Biomedicine, University of Palermo, Palermo, Italy.
[o] IRCCS Istituto delle Scienze Neurologiche di Bologna, Bologna, Italy.
[p] Istituto Italiano di Tecnologia, Nanoscopy, Genova, Italy.
[q] Institute of Biomedical Engineering, Department of Engineering Science, University of Oxford, UK.
[r] Hong Kong University of Science and Technology (Guangzhou), Guangzhou, China.
[s] INSA Rouen Normandie, Univ Rouen Normandie, Université Le Havre Normandie, France.
[t] Nuclear Medicine Department, Henri Becquerel Cancer Center, Rouen, France.
[u] Department of Radiology, Wake Forest University School of Medicine, USA.
[v] Department of Computer Science, University College London, United Kingdom
[w] School of Biomedical Engineering & Imaging Sciences, King's College London, UK

## Abstract

Airway-related quantitative imaging biomarkers are crucial for examination, diagnosis, and prognosis in pulmonary diseases. However, the manual delineation of airway structures remains prohibitively time-consuming. While significant efforts have been made towards enhancing automatic airway modelling, current public-available datasets predominantly concentrate on lung diseases with moderate morphological variations. The intricate honeycombing patterns present in the lung tissues of fibrotic lung disease patients exacerbate the challenges, often leading to various prediction errors. To address this issue, the 'Airway-Informed Quantitative CT Imaging Biomarker for Fibrotic Lung Disease 2023' (AIIB23) competition was organized in conjunction with the official 2023 International Conference on Medical Image Computing and Computer Assisted Intervention (MICCAI). The airway structures were meticulously annotated by three experienced radiologists. Competitors were encouraged to develop automatic airway segmentation models with high robustness and generalization abilities, followed by exploring the most correlated QIB of mortality prediction. A training set of 120 high-resolution computerised tomography (HRCT) scans were publicly released with expert annotations and mortality status. The online validation set incorporated 52 HRCT scans from patients with fibrotic lung disease and the offline test set included 140 cases from fibrosis and COVID-19 patients. The results have shown that the capacity of extracting airway trees from patients with fibrotic lung disease could be enhanced by introducing voxel-wise weighted general union loss and continuity loss. In addition to the competitive image biomarkers for mortality prediction, a strong airway-derived biomarker (Hazard ratio>1.5, p<0.0001) was revealed for survival prognostication compared with existing clinical measurements, clinician assessment and AI-based biomarkers.

* indicates co-last authors
Send correspondence to g.yang@imperial.ac.uk, y.nan20@imperial.ac.uk, x.xing@imperial.ac.uk

## 1. Introduction

Pulmonary fibrosis leads to symptoms such as dyspnoea and eventually results in mortality. Unfortunately, there is no cure for pulmonary fibrosis, and only a few therapeutic options are available to manage symptoms to slow its progression. Among all lung fibrosis, idiopathic pulmonary fibrosis (IPF) is the most serious chronic disease that scars the lung and makes it progressively more difficult for breathing, without a known cause (Lynch et al., 2018). However, invasive lung biopsy, which supports the golden standard of IPF, is barely accepted by patients with fibrosis due to its high risk (2% mortality rate in 100 days after biopsy) and cost (Han et al., 2015; Utz et al., 2001). As a result, there is an urgent need for clinicians to find a non-invasive clinical measurement for diagnosing IPF at its early stage. Previous work has identified a correlation between the severity of traction bronchiectasis on HRCT and the profusion of fibroblastic foci (areas of active collagen synthesis and reflect disease activity) on surgical lung biopsy samples taken from fibrotic lung disease patients (Walsh et al., 2015). Therefore, we hypothesize that the distention of the airways due to surrounding fibrosis (i.e., traction bronchiectasis) may be the earliest sign of evolving fibrosis. This kind of manual assessment requires airway modelling, which is tedious and unaffordable with nearly 5 hours annotated from scratch. In addition to the heavy workload, visual assessment from experts is also subjective and may ignore subtle changes in airway dimensions over short periods, especially in the early stage when there are no extensive fibrotic patterns. Compared to the manual assessment, which is subjective, imprecise, poorly reproducible, and time-consuming, automatic qualification of airways can achieve objective and stable performance at a lower cost.

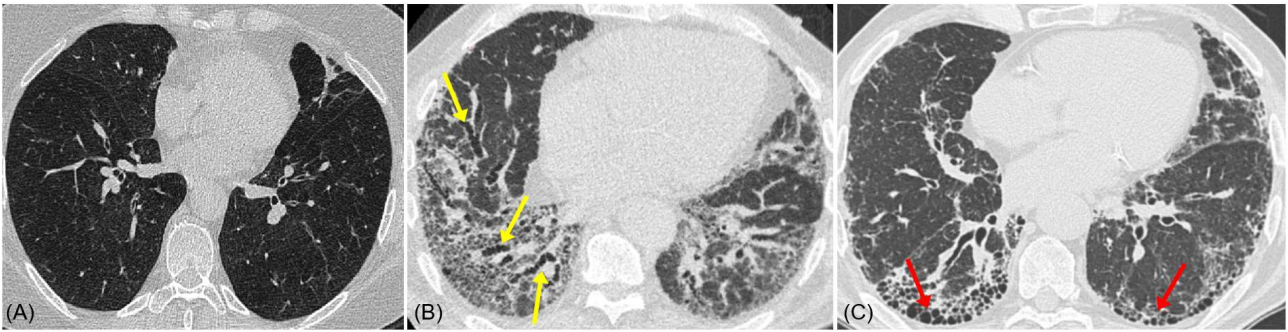

Fig. 1. HRCT collected from (a) patients with nodule (LIDC dataset), (2) patients with traction bronchiectasis (yellow arrow), and (3) patients with honeycombing (red arrow). The honeycombing makes the airway trees hard to be extracted and leads to poor airway modelling capabilities.

Although several challenges and publicly available datasets have been released for airway modelling, limited attention has been paid to lung fibrosis. For instance, the existing open challenges, EXACT'09 (Lo et al., 2012) and ATM22 (Zhang et al., 2023), published cases collected from healthy volunteers and patients with common lung diseases (e.g., patients with nodes, asthma, etc.). Although ATM22 included some COVID-19 cases for additional assessment, there is still no publicly available data for complex fibrotic lung disease. Different from other pulmonary diseases, patients with fibrosis suffer from volume loss, bronchiectasis, distortion of the secondary lobule, and the loss of the smooth air-to-interstitium interfaces. These abnormal changes, especially the bronchiectasis, lead to distinct airway changes compared with other pulmonary diseases. Different from patients with other lung diseases (Fig. 1(a)), patients with fibrosis suffer from bronchiectasis, with significant distension of the terminal/small branches in Fig. 1(b). Moreover, honeycombing, one of the UIP-like (Usual Interstitial Pneumonia) patterns in fibrosis cases, exacerbates the difficulty of automatic airway modelling and makes the computational model error-prone. Specifically, honeycombing refers to the presence of small cystic airspaces in the lung periphery, which shows similar patterns to the bronchus (Figure. 1 (c)). The honeycomb within the lung tissue in patients with fibrotic lung disease leads to heavy airway leakages and mis-predictions.

The main challenge in developing reliable AI systems for fibrotic lung disease is the abnormal morphology changes of lung tissues and airway trees, which include 1) **Airway leakages.** The airway leakages were initially mentioned in (Lo et al., 2012), referring to the number of false-positive predictions (voxel). Technically, this issue is mainly due to the similar patterns between airway trees (especially for terminal branches) and lung periphery tissues. For patients with fibrosis, the honeycombing intensively confused the AI model. Specifically, these honeycombs within the lung periphery were regarded as airway branches, leading to a large number of airway leakages. Although the EXACT'09 organizers reported the leakage counts and leakage volumes, they cannot well assess the leakage conditions since the scale of lung tissues varies for individuals. 2) **Missing and discontinuity airways.** Due to the various imaging conditions, the airway intensity value may be close to that of the lung tissues. Under such conditions, AI models are prone to false negative predictions and take the airway voxels as background regions. 3) **Weak robustness in fibrosis cases.** Airway models that developed on normal lung diseases cannot present high robustness in fibrosis cases, due to the complex abnormal changes of the lung tissue and airway trees. According to an existing study (Nan et al., 2023), models trained on normal lung disease demonstrated a significant performance gap in fibrosis cases compared with that on COVID-19. Interestingly, all the models performed well in COVID cases with no significant decrease in detected branch ratio and length ratio. 4) **Limited research on AI prognostication in fibrosis cases.** Although studies have shown manual-derived biomarkers have an association with mortality (Best et al., 2008), it remains unclear whether the automatically airway-derived imaging biomarkers can be used for prognostication.

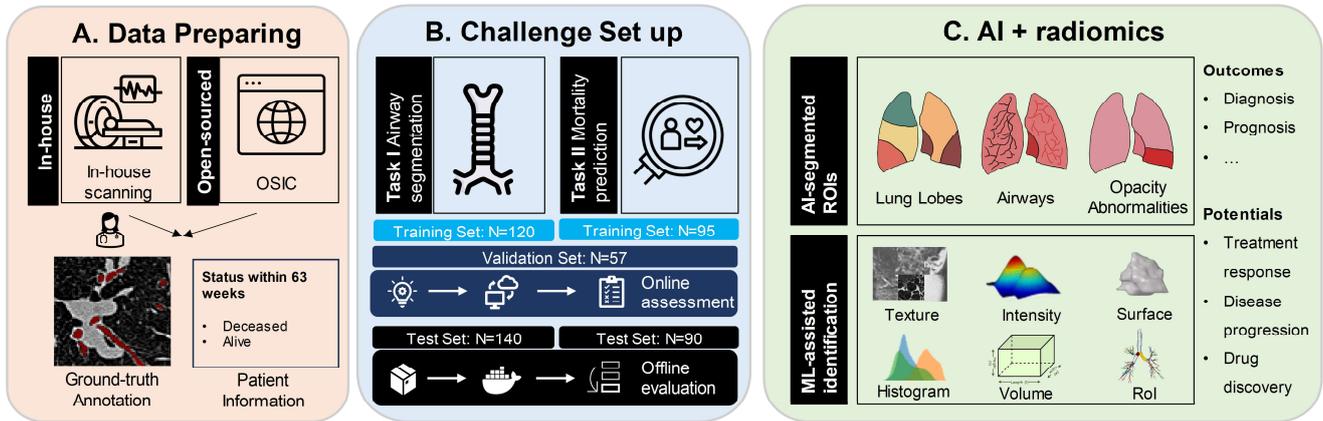

Fig. 2. The workflow of AIIB23 challenge. A) high-resolution CT scans (HRCT) were collected and annotated by junior and senior radiologists. B) the challenge included two subtasks, the airway segmentation and the mortality prediction. Two assessment stages were included, with the online assessment that received the prediction from participants and offline assessment which used docker images for evaluation. C) the imaging biomarkers (shape, textural, statistical features including airway indicators) could be used for clinical prognosis and treatments. RoI refers to region of interests.

The 2023 Airway-Informed Quantitative CT Imaging Biomarker for Fibrotic Lung Disease Challenge (AIIB23) in conjunction with the MICCAI-2023 (Fig. 2) aimed to (1) accelerate and assess the progress on the advanced computational approaches for complex airway extraction, (2) investigate the imaging biomarkers for prognostication in clinical practice, and (3) demonstrate the research directions and existing challenges. Hosted on the CodaLab platform (Pavao et al., 2022), AIIB23 was comprised of two subtasks, including Task I for airway segmentation and Task II for mortality prediction. We received 148 team registrations from more than 300 participants for the challenge. The official leaderboard indicates the performance of submissions from teams who meet all the criteria. Specifically, participants from 20 teams were ranked based on overlapped metrics and branch scores across 52 cases, and 10 teams were finally ranked by assessing both accuracy and efficiency across the 140 offline test cases. The contributions of this paper can be organised as follows:

- AIIB23, as the first open challenge and publicly available dataset for lung fibrosis disease, builds a new benchmark of AI-based solutions for airway modelling on various datasets. The released data include 120 HRCT scans with rare expert annotations, which is the first airway reference on fibrotic lung disease.

- AIIB23 incorporates a prognostication task (mortality prediction) on fibrotic lung disease. Participants are required to develop sophisticated models that are not only for airway extraction but also for investigating the potential progression of fibrosis, ensuring a more holistic and forward-thinking

- AIIB23 demonstrates the state-of-the-art airway modelling approaches and highlights the remained challenges for current AI models in airway segmentation. AIIB23 investigates independent airway-derived biomarkers for mortality predictions of patients with fibrotic lung disease.

The paper is organized as follows: In Section 2, we first introduce the related works in airway segmentation, AI-based prognostication in fibrotic lung disease, and publicly available datasets for airway segmentation. Section 3 goes deeper into the AIIB23 challenges, introducing the details as well as the criteria for airway segmentation and mortality prediction tasks. Section 4 reports the results of AIIB23 challenges, and further analyses the collected results via additional offline evaluations. The BIAS checklist is demonstrated in the supplementary files referring to the guidance in (Maier-Hein et al., 2020).

## 2. Related work

### 2.1. Airway segmentation

Airway segmentation is the essential step for airway-derived imaging biomarkers, which refers to modelling the airway trees to three-dimensional (3D) binary masks. Given an input HRCT scan $X$, the airway segmentation model $\Phi$ aims to extract the masks of airway trees $M_A = \Phi(X)$. Efforts have been made on various pulmonary diseases, such as lung cancers (Qin et al., 2019; Zheng et al., 2021) and common pulmonary diseases (Charbonnier et al., 2017; Graham et al., 2010; Lo et al., 2012; Xu et al., 2015) (e.g., lung nodes, ground glass, infiltrates). To our knowledge, the first work that focused on airway segmentation was in 1994, in which Sonka et al. proposed a knowledge-based method to segment intrathoracic airway trees (Sonka et al., 1996). Subsequently, researchers applied image processing methods such as morphology operations, region growing (Tschirren et al., 2005), intensity thresholds (Aykac et al., 2003), and fuzzy connectedness (Xu et al., 2015) to extract airway trees. For instance, Xu et al. proposed a fuzzy affinity relationship to modify the fuzzy connectedness segmentation and used a spatially constrained Markov random walk for estimating the thickness of airway walls (Xu et al., 2015). Mori et al. applied the region growing method to extract the bronchus region from the 3D CT images and used the centreline for anatomical labelling (Mori et al., 2000). However, these conventional approaches were reported with weak robustness and reproducibility, due to the variations of airway voxel intensities (Li et al., 2022).

Starting in 2015, deep learning has dominated the computer vision field and many researchers have implemented deep learning-based semantic segmentation for airway extraction. Charbonnier et al. implemented a 2D convolutional neural network (CNN) to reduce airway leakages by classifying candidate airway patches (given by conventional approaches) into true-positive or false-positive predictions (Yun et al., 2019). Yun et al. presented a 2.5D CNN that incorporates more spatial information by taking three orthogonal directions 2D slices as inputs (Yun et al., 2019). In addition to 2/2.5-D CNN, 3D CNN outperformed these methods by incorporating more spatial information. To address the shortage of GPU memory when training on 3D volumetric data, researchers proposed the patch-based strategy to reduce memory consumption, as well as cropping the lung region out of the original HRCT scans (Garcia-Uceda Juarez et al., 2018; Qin et al., 2019). The commonly used baseline model for 3D airway segmentation is the 3D U-Net, which comprises skip connections, an encoder, and a decoder. Among various U-Net variants, nnUNet (Isensee et al., 2021) was known as the most robust and reproducible approach and was widely used in the previous challenge (Zhang et al., 2023). Additionally, researchers also proposed other techniques such as graph refinement (Jin et al., 2017), connectivity awareness (Qin et al., 2019), centreline-aided learning (Reynisson et al., 2015), hard samples digging (Zheng et al., 2021), fuzzy attention mechanism (Nan et al., 2023), and transformer-based 3D U-Net (Wu et al., 2023). For instance, Nadeem et.al, (Nadeem et al., 2020) and Qin et al., (Qin et al., 2020) modified the vanilla 3D U-Net for airway segmentation by implementing voxel-connectivity awareness and freeze-and-grow propagation to alleviate the airway leakages. Existing approaches have been noted to struggle with fibrosis cases with complex abnormalities like bronchus deformation, ground glass/consolidation, and honeycombing, even when trained on numerous cases of other pulmonary diseases (Nan et al., 2023). For instance, the test performance of fibrosis cases presents a significant gap compared with that of the COVID-19 cases, with approximately 8-14% descent of intersection of union (IoU) score, 15-30% drop of detected length ratio (DLR), and 17-30% decline of detected branch ratio (DBR) (Nan et al., 2023). However, it remains unclear how these models perform on fibrosis cases when trained specifically on them, and how these models performs when they were further tested on COVID-19 and other pulmonary diseases.

## 2.2. AI-driven prognostication of lung fibrosis

Radiology is a vital part of diagnosis and disease management in pulmonary disease, especially for patients with lung fibrosis. A chest CT showing typical features/patterns can be used for the diagnosis of idiopathic pulmonary fibrosis (IPF) without pathologic confirmation (Raghu et al., 2018). Specifically, the American Thoracic Society/European Respiratory Society/Japanese Respiratory Society/Latin American Thoracic Association (ATS/ERS/JRS/ALAT) guidelines have included certain imaging patterns in CT scans (e.g., honeycombing) as the key prognostic factors (Raghu et al., 2011). However, accurately quantifying pulmonary fibrosis from CT scans remains challenging, limiting its current prognostic use. As a result, even with certain disadvantages, physiologic indicators such as FVC and CPI are still widely used as the primary prognostic biomarkers.

To address these issues, researchers have investigated the prognostic values of imaging-derived biomarkers. The first step to HRCT-derived biomarkers for the prognosis of lung fibrosis was proposed by Best et al. (Best et al., 2003), indicating a correlation between kurtosis of the lung HRCT histogram and mortality. CALIPER, known as computer-aided lung informatics for pathology evaluation and ratings, was another AI-driven method for lung fibrosis prognosis. CALIPER uses histogram characteristics to classify and quantify the disease and can predict outcomes in various interstitial lung diseases (ILDs) (Jacob et al., 2016). In addition, Nam et al., developed a cascaded AI model (comprising two U-Nets) to segment honeycombing (Nam et al., 2023), consolidation, ground glass opacity, reticular opacity, and emphysema, demonstrating that the normal lung volume proportion and fibrotic lung volume were independent prognostic factors for overall survival in IPF. Humphries et al., (Humphries et al., 2018) proposed an HRCT-derived measurement for IPF diagnosis and prognosis, exploring the correlation between image-derived biomarkers (quantitative measurement of fibrosis regions) and physiological measurements such as forced vital capacity (FVC), diffusing capacity of the lung for carbon monoxide (DLCO). The results indicated that the minimal clinically important difference of HRCT-derived biomarkers was 3.4%. Although efforts have been made, it is unclear whether airway-derived biomarkers can be used for the prognosis of lung fibrosis diseases.

## 2.3 Publicly available datasets and open challenges

This section demonstrates the existing publicly available datasets and open challenges associated with airway segmentation and lung fibrosis prognostication, which aims to highlight the differences between these events and AIIB23.

**EXACT'09:** EXACT'09 (Lo et al., 2012) is the first open challenge for airway segmentation, which includes 40 cases to evaluate 15 different airway segmentation algorithms. These cases were collected from eight institutions under different acquisition protocols and various devices, either at peak inspiration or full expiration, showcasing a range of pathological abnormalities. The 40 cases were divided into training and test sets, with 20 for each, respectively. It is of note that only the training cases were released to the public. The slice sickness of BAS ranges from [0.6, 1.25] and the reconstruction kernels include B50f, FC12, FC10, B30f and B70f. The anomalies include healthy volunteers, patients with small lung nodules, ground glass, diffuse emphysema, etc. Methods developed on EXACT'09 mainly included image processing approaches, e.g., using morphology approaches (Irving et al., 2009), and region-growing variations (Mendoza et al., 2009; Pinho et al., 2009).

**ISICDM21:** ISICDM21 includes 12 cases with expert annotations, the size of each slice is 512*512 and the slice thickness ranges from 0.625mm to 1.5mm (Wu et al., 2023). Unfortunately, this dataset is not accessible currently.

**BAS:** Binary airway segmentation (BAS) (Zheng et al., 2021) is a publicly available dataset with 90 HRCT cases. Among the 90 cases, 20 cases were obtained from EXACT'09, and the other 70 cases were collected from the lung image database consortium image collection (LIDC) (Armato III et al., 2011). The slice sickness of these 70 cases ranges from 0.45 to 0.625 mm, and the abnormalities mainly include lung nodules. The number of slices of BAS data ranges from 157 to 764, indicating various scanning protocols.

**ATM22**: ATM22 (Zhang et al., 2023) is a multicentre airway segmentation dataset with 500 cases in total (300 for training, 50 for validation, and 150 for testing). It released 300 cases with available semantic annotations to the public. Among these 300 cases, 90 cases were obtained from the BAS dataset, while 210 cases were collected from LIDC-IDRI data and Shanghai Chest Hospital. The number of slices ranges from 157 to 1125, with the slice thickness ranging from 0.45mm to 1.00mm.

## 3. Materials and methods

### 3.1. The AIIB23 Dataset

AIIB23 integrates lung CT scans from patients with fibrosis and COVID-19 collected from OSIC, AIPFR and UHP (demographics shown in Table 1), using different vendors or scanning protocols. Data from the OSIC repository were randomly split into training and validation sets and released to the public, while those from AIPFR and UPH were used for the offline test set (Fig. 2). All registered participants agreed to the CC-BY-NC-ND license.

- **OSIC**[1] is an open-access global dataset that includes multidimensional medical images (e.g., 3D HRCT, 3D MRI) of IPF, fibrotic interstitial lung disease (ILD) and other respiratory diseases (e.g., emphysematous diseases) with clinical metadata including physiologic indicators (e.g., FVC percentage and DLCO), body measurements (e.g., height, age, and gender), mortality status and clinical diagnosis. The database was reviewed by two global GDPR / HIPAA privacy firms to ensure the maximum size of the dataset and diversity of data.

Table 1. Demographics of data used in AIIB23 challenge.

| **Attribute** | OSIC (n=172) | AUS (n=90) | UHP (n=50) |
|---|---|---|---|
| Age (years) | 68 (57, 75) | 70 (67, 76) | 65 (45, 82) |
| BMI (kg/$m^3$) | - | 28.39 (26.27, 30.92) | - |
| **Gender** | | | |
| Male | 101 (58.7%) | 54 (61.1%) | 24 (48%) |
| Female | 71 (41.3%) | 36 (38.9%) | 26 (52%) |
| **Diagnosis** | | | |
| ILD | 99 (57.6%) | 0 | 0 |
| IPF | 62 (36.0%) | 71 (78.9%) | 0 |
| COVID-19 | 0 | 0 | 44 (88%) |
| Other abnormalities | 11 (6.4%) | 19 (21.1%) | 6 (12%) |

\* Continuous variables such as age and BMI are presented as Median (Q1, Q3).

- **AIPFR**. The Australian IPF repository (AIPFR) (Walsh et al., 2020) has ethical approval from the Sydney Local Health District (protocol no. X14-0264). The diagnosis for each patient includes IPF, probable IPF, alternative diagnosis, and other fibrotic patterns based on the 2018 ATS/ERS/JRS/ALAT IPF guideline statement (Raghu et al., 2018).
- **UHP.** The UHP data follows the approval of the local Ethics Committee (code 934/2021/OSS/AOUPR - 11.01.2022) at the University Hospital of Parma (UHP). All patients included were admitted to UHP with symptoms and/or signs highly suggestive of COVID-19, with PCR test results.

All HRCT scans were initially collected in DICOM format and converted to Nifti format via SimpleITK 2.1.0 package in Python 3.10.0. The preliminary annotation was first given by an AI model trained on the BAS dataset (Nan et al., 2023), followed by careful revision and modifications of three junior radiologists. These annotations were then revised and refined by two senior radiologists by delineating the breakages and mis-annotations of airway voxels. We followed the annotation given by most experienced radiologists when there are conflicting opinions between radiologists regarding a certain case. The annotations were conducted based on the drawing pad (HUION Kamvas 16 Tablet) via ITK-SNAP 4.0.1 (Yushkevich et al., 2006). In particular, the inter-observer agreement between radiologists for airway branches was almost perfect (Cohen's Kappa=0.937). The challenge data was randomly chosen from the three databases, ensuring only high-quality cases were retained. Specifically, we selected cases where the number of slices surpassed 120 and each slice measured over 512x512 pixels. Overall, 120 (95) cases were set as training data for Task I (Task II), 52 cases were set as a validation set for both Task I and II, and 140 (90 fibrosis and 50 COVID-19) cases were set as test set (Fig. 2). All the cases were converted to NIFTI format before releasing to the public.

AIIB23 presents diverse imaging qualities, with the differences mostly being attributed to the number of slices. To better explore the data distribution, we summarized the slice number for each case in training, validation and test sets, respectively (Fig. 3. A). The results indicated that the training and validation data (OSIC) possess a higher quality compared to the test fibrosis data (AIPFR), as evidenced by a more significant number of slices ($p<0.05$). Specifically, the image quality of the coronal and sagittal planes in low-slice CT scans (Fig. 3. B. 5 and 6) is weaker than that in high-slice CT scans (Fig. 3. B. 2 and 3), with more visually blurry and jagged phenomenon. Additionally, this also affects the expert annotations due to the loss of fine-grained information between slices.

---
[1] https://www.osicild.org

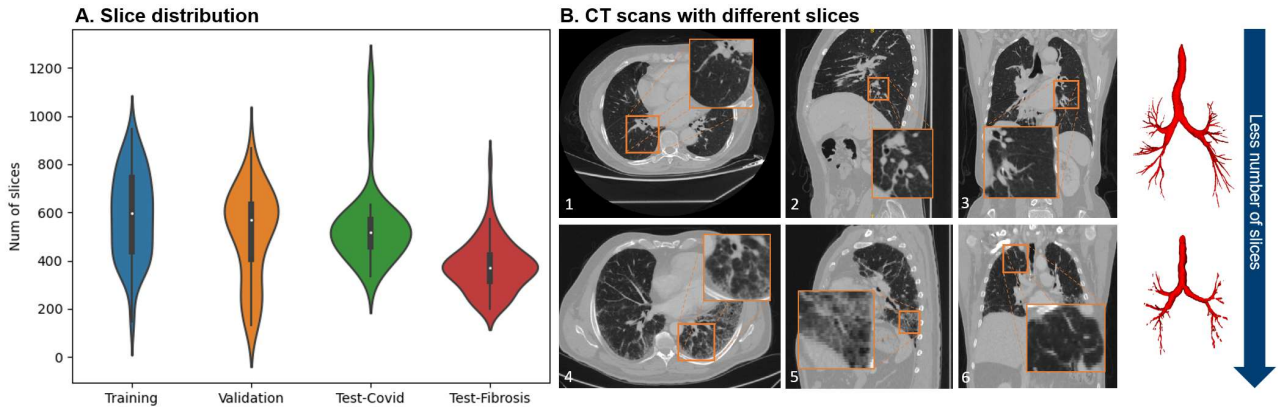

Fig. 3. Variation in the number of slices of the AIIB datasets. A) the violin plot of the number of slices for each case in training, validation, and two test sets, respectively. B) Visualization of two CT scans (with different slices) and their corresponding annotations (B.1-B.6). First row: a CT scan with 1145 slices; second row: a CT scan with 140 slices.

### 3.2. Challenge setup

The AIIB23 challenge was divided into two subtasks, Task I-airway segmentation, and Task II-mortality prediction. Task I encouraged participants to segment airway trees from the HRCT scans, specifically focusing on the extraction of small airway branches. Task II asked participants to predict the mortality status (survival or not) of patients at a certain time step (63 weeks after the first HRCT scanning and diagnosis date).

### 3.2.1. Schedules

The data was annotated between January 2023 and April 2023, after the challenge proposal was submitted to MICCAI 2023. There were two tasks for AIIB23, one was the airway segmentation task, and the other was the mortality prediction task. Details of these two tasks are given in Figure. 2. The accept notification was received on 3$^{rd}$ March 2023 and the official challenge page was opened[2,3] on 15th May (on the CodaLab platform), with the challenge registration opened and training data released. The validation cases were released by the end of May (30$^{th}$) and the online validation submissions were allowed on the same day. On 30$^{th}$ July, the test phase was opened, and participants were required to submit their docker files by 25$^{th}$ August. The top ten and top five methods of Task I and Task II were publicly announced. It is of note that this challenge is an open challenge and will continually serve as a benchmark for AI in fibrotic lung disease.

### 3.2.2. Highlights

**Reproducibility**. In the test phase, teams were required to submit a docker image with a short paper describing their approaches. The scores they obtained during the validation phase will be evaluated using the submitted docker image. In addition, participants were encouraged to make their original code publicly available (see details in [4]).

**Fairness.** No external data is allowed when training the model, while it is feasible to use pretrained weights (not on airway-related datasets) as pre-training and domain-adaptation approaches may be applied. The test environment for all competitors remains the same, with 1 GTX3090 TI (24GB) and 64 GB RAM provided. Specifically, participants from the organizers' research lab were not allowed to register for the challenge, and those from the same institute could participate but were ineligible for awards.

**Submission restriction.** All participants were required to apply for the challenge using their official email address to prevent the same team from submitting multiple times with different accounts. Each team can only have 50 times submissions during the validation phase to prevent the leakage of ground truth labels by enormous random trials. During the test phase, each team can only have one successful submission (the successful submission indicates the docker image can be successfully executed and can produce predictions). The instructions for submission were shown on the challenge web page[5].

---
[2] https://codalab.lisn.upsaclay.fr/competitions/13238
[3] https://codalab.lisn.upsaclay.fr/competitions/13256
[4] https://codalab.lisn.upsaclay.fr/competitions/13238#learn_the_details-final-results

**3.2.3. Evaluation metrics**

The evaluation metrics for AIIB23 can be summarized as two parts corresponding to two subtasks.

- **Task I-Airway segmentation**. Task I applies both overlapped measurements and branch measurements to assess the model performance. Assume $X$ as the predictions and $Y$ as the manually annotated labels, metrics for Task I mainly include intersection over union (IoU) score, detected length ratio (DLR), detected branch ratio (DBR), precision, airway leakage ratio (ALR), and inference time, which are given as

$$\text{IoU} = \frac{X \cap Y}{X + Y}, \quad (1)$$

$$\text{DLR} = \frac{L_X}{L_Y}, \quad (2)$$

where $L_X$ and $L_Y$ indicate the branch length for prediction $X$ and ground truth $Y$, respectively.

$$\text{DBR} = \frac{N_X}{N_Y}, \quad (3)$$

$$\text{Precision} = \frac{TP}{TP + FP}, \quad (4)$$

$$\text{ALR} = \frac{FP}{TP + FN}, \quad (5)$$

where TP is the number of true positive predicted voxels, FP is the number of false positive predicted voxels, FN is the number of false negative predicted voxels and $N_X$ is the number of correctly detected branches and $N_Y$ is the number of branches in the ground truth. The successful detection of a branch is defined as that branch has a high IoU score (threshold=0.8) with a certain ground truth branch. The final results were ranked based on both the overall accuracy (OvAcc) of airway modelling and the efficiency of the algorithms (inference time $t$), with

$$\text{OvAcc} = (\text{IoU} + \text{Precision} + \text{DBR} + \text{DLR}) * 0.25, \quad (6)$$

$$R = \mathcal{R}[\text{OvAcc}] * 0.7 + \mathcal{R}(t) * 0.3, \quad (7)$$

where $\mathcal{R}$ is the ranking operation $\mathcal{R}(x) \in \{0,1,2,\ldots N\}$ and $t$ indicates the inference time (seconds).

- **Task II-mortality prediction.** The prognosis task is structured as a binary classification problem aimed at predicting the mortality status of patients after 63 weeks from their initial scan, where 0 denotes deceased, and 1 denotes alive. The numbers of alive/deceased cases for train-validation-test splits were 59/36, 26/26, and 78/12, respectively. The evaluation metric is a composite that includes Accuracy, AUC (Area Under the Curve), Sensitivity, Specificity, and F1 Score. These metrics provide a comprehensive evaluation of the model's performance in predicting patient mortality after 63 weeks, considering both the ability to correctly classify deceased and alive patients and the trade-off between sensitivity and specificity. The time threshold (63 weeks) was grounded in an empirical evaluation of our dataset, aiming to optimize the balance between the number of deceased cases included in the study and the follow-up period. A model with high specificity can accurately identify high-risk patients (who will decease within 63 weeks).

- **Missing or collapse results.** The assessment algorithm would be terminated if there were collapse or missing results in the prediction of submissions. Specifically, competitors could obtain the log file to locate the cases that failed to be predicted and revise their algorithms.

- **Statistical analysis.** The Wilcoxon signed-rank test was used to analyse the significance between two different comparisons, and significance was observed when p-value<0.05. The Cox proportional hazard modelling was used for survival analysis to determine crude and adjusted hazard ratios.

**3.3 Participants**

As an official MICCAI 2023 challenge, AIIB23 received 214 applications for Task I and Task II registration, with 112 for Task I and 102 for Task II, respectively. Among all these applications, 53 applications were rejected according to the terms and conditions. We received validation results from 20 teams of Task I and 8 teams of Task II (details are presented in S-Table 1). The top 10 approaches for Task I and the top 5 methods for Task II have been included in this paper, with participants consenting to the inclusion of their solutions and results in the publication.

---
[5] https://codalab.lisn.upsaclay.fr/competitions/13238#learn_the_details-submission-guidance

## 4. Results

The detailed description of methodologies submitted by competitors of Task I and Task II is demonstrated in the supplementary file (section named 'Solutions in AIIB23').

### 4.1. Performance of airway modelling

Methodologies proposed by different teams in airway modelling shared certain common points (Table. 2). Initially, all the submitted models were deep learning-based solutions, and the comprehensive methodologies of each team

Table 2. Summary of solutions proposed by top-10 teams in Task I.

| Teams | Preprocessing | | | | Data sampling | DA | ME | Loss |
|---|---|---|---|---|---|---|---|---|
| | WindowT | RoIEx | SpacingT | Norm | | | | |
| MedibotTeam | ✗ | ✓ | ✓Median | ✓z-score ✓max-min | Random crop Hard-mining | ✓ | ✓ | $L_G, L_{Dice}, L_C$ |
| IMR | ✗ | ✓ | ⚔ Median | ✓max-min | Targeting crop | ✓ | ✓ | $L_G, L_{CE}, L_{CA}$ |
| Infervision | [-1000,500] | ✓ | ✗ | ✓max-min | Random crop Hard-mining Targeting crop | ✓ | ✗ | $L_G, L_{Dice}$ |
| Sanmed_AI | [-1400, 200] | ✓ | ✗ | ✓z-score | Crop | ✓ | ✗ | $L_{Dice}, L_{FL}$ |
| Gexing | ✗ | ✗ | ✓Median | ✓z-score | Random crop | ✓ | ✓ | $L_{Dice}, L_{CE}$ |
| DJ_92 | ✗ | ✗ | ✓Median | ✓z-score | Crop | ✓ | ✗ | $L_{Dice}, L_{CE}$ |
| Riipl | [-1400, 600] | ✗ | ✓Median | ✓z-score | Random crop | ✓ | ✓ | $L_{Dice}, L_{CE}$ |
| earth1is1flatten | ✗ | ✗ | ✗ | ✓z-score | Random Crop | ✓ | ✗ | $L_{Dice}$ |
| Dolphins | ✗ | ✓ | ✓Median | ✓z-score | Random Crop | ✓ | ✗ | $L_{Dice}, L_{CE}$ |
| Junqiangmler | [-1000,-300] | ✓ | ✓Mean | ✓z-score | Random Crop | ✓ | ✗ | $L_{Dice}, L_{CE}$ |

Notes: ⚔ indicates at least one of the models included this technique, WindowT: window transform (if yes, presenting the HU range), RoIEx: RoI (lung) extraction, SpacingT: spacing transform, Norm: normalization, DA: data augmentation, ME: model ensembling, $L_G$: general union loss, $L_{Dice}$: dice coefficient loss, $L_C$: airway continuity loss, $L_{CA}$: connectivity-aware loss, $L_{CE}$: cross entropy loss, $L_{FL}$: focal loss.

can be found in the Supplementary material.

All methods were formulated using a CNN-based patch-learning strategy due to the limit of GPU memories. This approach involves training CNN-based models on sampled patches while combining the predictions from all patches during the inference phase. In addition, more than half of the competitors (6/10) performed lung extraction before data sampling to reduce the artefacts outside the lung region. This includes applying open-source lung extraction algorithms (Hofmanninger et al., 2020) and utilizing a support vector machine (SVM) to classify different connected components. The majority of participants applied spacing transform to alleviate the anisotropy of raw CT data, which was achieved by resampling the voxel spacing to the median spacing of the entire training set. Interestingly, only four teams performed window transforms to the raw CT data, while others rescaled the HU values by introducing normalization strategies (e.g., max-min or z-score normalization). All top-3 methods employed weighted general union loss $L_G$ (Zheng et al., 2021) and connectivity loss (e.g., $L_C$ and $L_{CA}$) (Nan et al., 2023), with multi-stage training protocols. In addition to the commonly used random cropping for data sampling, all competitive methods (top-3 approaches) applied specific data sampling strategies, such as hard mining (sampling patches according to the mispredicted airway skeletons) and targeting cropping (sampling patches around the airway skeletons). These data sampling strategies can better select samples around the RoIs and can effectively alleviate the data imbalanced issue. In particular, model ensembling also leads to better airway modelling performance, however, will also decrease the time cost if no multiple threads techniques are implemented.

The overall performance of the top-10 teams is demonstrated in Table 3, ranked by the overall accuracy of the model and the inference time cost. MedibotTeam achieved the best results in the test phase considering both inference speed (43.79 s/scan) and model accuracy (0.9185), which is 30.4% faster than the IMR team (62.98s/scan). The IMR team gained the best branch metrics among all the competitors, with the 0.9510 detected length ratio (DLR) and 0.9312 detected branch ratio (DBR). However, their model also had some weaknesses, with a high airway leakage ratio (ALR) of 0.1089 and a 0.9014 precision score. Teams Gexing achieved the highest overlapped metric with a 0.9087 IoU score. Unfortunately, their method over-focused on the IoU score and achieved relatively low branch scores (0.8603 DLR and 0.7765 DBR) compared with top-performing teams. The model proposed by team DJ_92 achieved the highest precision (0.9619) and the lowest ALR (0.0397), indicating that the model may mis-segment some terminal branches which contribute a little to the overlapped metrics. Interestingly, although team dolphins achieved a comparable overall accuracy (0.8597), the time consumption is huge taking nearly 1500s for each CT scan.

Table 3. Top-10 winners in AIIB23 Task I (n=140).

| Teams | IoU | DLR | DBR | Precision | ALR | AMR | OvAcc | Time (s/per scan) |
|---|---|---|---|---|---|---|---|---|
| MedibotTeam | 0.9049 | 0.9365 | 0.9051 | 0.9276 | 0.0786 | **0.0259** | **0.9185** | **43.79** |
| IMR | 0.8770† | **0.9510†** | **0.9312†** | 0.9014† | 0.1089† | 0.0299† | 0.9152† | 62.98† |
| Infervision | 0.9016† | 0.9201† | 0.8825† | 0.9399† | 0.0639† | 0.0425† | 0.9110† | 87.93† |
| Sanmed_AI | 0.8903† | 0.8838† | 0.8354† | 0.9203† | 0.0854† | 0.0404† | 0.8825† | 92.78† |
| Gexing | **0.9087†** | 0.8603† | 0.7765† | 0.9518† | 0.0499† | 0.0466† | 0.8743† | 356.04† |
| DJ_92 | 0.8967† | 0.8466† | 0.7504† | **0.9619†** | **0.0397†** | 0.0686† | 0.8639† | 223.66† |
| Riipl | 0.9003† | 0.8560† | 0.7672† | 0.9584† | 0.0423† | 0.0620† | 0.8705† | 531.52† |
| earth1is1flatten | 0.8221† | 0.6986† | 0.6043† | 0.9325† | 0.0906† | 0.1538† | 0.7644† | 155.36† |
| dolphins | 0.8838† | 0.8556† | 0.7627† | 0.9365† | 0.0682† | 0.0580† | 0.8597† | 1500.12† |
| Junqiangmler | 0.7488† | 0.7853† | 0.6853† | 0.8200† | 0.2188† | 0.1313† | 0.7599† | 316.89† |

† indicates significant differences compared with the top-1 result (p<0.05).

In addition to the overall performance, the model capacities on COVID-19 and fibrosis cases were reported in S-Table 2 and S-Table 3, respectively. In particular, all models achieved better performance when testing on out-of-distribution (OOD) data (COVID-19 cases), with a nearly 1-2% increase in IoU scores. Specifically, there witnessed a sharp increase in branch metric scores for models proposed by DJ_92 (11.7% DLR and 16.1% DBR), earth1is1flatten (16.8% DLR and 18.4% DBR), and Riipl (12.3% DLR and 17.3% DBR). This indicated that models trained on fibrosis cases can be well implemented in COVID-19 cases.

### 4.2. Performance of mortality prediction

In mortality prediction studies, two distinct strategies have been observed. The first integrated radiomics with machine learning, where Pyradiomics was used for extracting radiomics features. For instance, the uAI team

Table 4. Summary of solutions proposed by top-5 teams in Task II.

| Teams | CT Windowing | Method Type | Semantic input | ME | Classifier |
|---|---|---|---|---|---|
| uAI-Team | [-1350, 150] | ROI-based radiomics | ✓ lung lobes ✓ pneumonia | ✓ | LGBM, LR, SVM, and XGBoost |
| Tastefish | [-1000, 400] | End-to-end CNN | ✗ | ✗ | 3D CNN |
| junqiangmler | [-1000, 800] | End-to-end CNN | ✓ lung region | ✗ | 3D ResNet16 |
| Earthis1flatten1 | ✗ | Image-based radiomics | ✓ airways | ✗ | XGBoost |
| DJ_92 | ✗ | ROI-based radiomics | ✓ trachea | ✗ | SVM |

Notes: LGBM: light gradient-boosting machine, LR: logistic regression, SVM: support vector machine, XGBoost: extreme gradient boosting, ME: model ensembling.

expanded this by adding hand-crafted features such as lesion volume and surface details. Machine learning models, e.g., XGBoost and SVM, are then applied to these features for binary classification, bypassing the extensive training typically required for convolutional neural networks. Conversely, another strategy was based on CNNs, notable for their omission of manually engineered features and enhanced speed in delivering predictions. In these solutions, 3D neural networks were employed and trained directly on the entire volumetric HRCT to perform binary classification tasks.

Table 5. Top-5 winners in AIIB23 Task II (n=90).

| Teams | Overall score | ACC | Sensitivity | Specificity | F1 score |
|---|---|---|---|---|---|
| uAI-Team | **0.7059** | **0.7528** | **0.7848** | 0.5000 | **0.8493** |
| Tastefish | 0.7049† | 0.6404† | 0.6203† | 0.8000† | 0.7538† |
| junqiangmler | 0.6222† | 0.5955† | 0.5949† | 0.6000† | 0.7231† |
| earth1is1flatten | 0.6186† | 0.6629† | 0.6962† | 0.4000† | 0.7857† |
| DJ_92 | 0.5787† | 0.4045† | 0.3291† | **1.0000**† | 0.4952† |

† indicates significant differences compared with the top-1 result ($p<0.05$).

Table 5 presents the performance of the top five teams in the AIlB23 Task II. The uAI-Team leads with the highest overall score of 0.7059, accompanied by a notable F1 score of 0.8493. Tastefish closely trails in the second spot, distinguished by a high specificity of 0.8000. Junqiangmler and earth1isflatten secure the third and fourth positions respectively, with the latter team boasting the second-highest F1 score at 0.7857. Interestingly, DJ_92, despite ranking fifth, stands out with a perfect specificity score of 1.0000 but low accuracy (0.4045) and sensitivity (0.3291). These rankings highlight the diverse strengths of each model across various evaluation metrics.

## 5. Discussion

In this section, we mainly discuss the research findings and conduct further analysis regarding the AIIB23 challenges. The summaries of common strategies used in airway modelling and mortality prediction were first introduced, followed by the additional analysis of the prediction errors of different teams. Moreover, we also discover the capabilities of different models against unharmonised data for clinical practice. In addition, reliant radiomic features were investigated by performing survival analysis. Last but not least, the survival analysis is performed based on the biomarkers proposed by challenge participants, existing studies, and our solutions.

**Overall performance of airway modelling**

The overall performance of Task I was varied but with interesting research findings. In particular, there were no significant differences of the OvAcc among top-2 teams, with 0.9185, and 0.9152 ($p>0.05$), respectively. This indicates that these models performed similarly in comprehensive evaluations but with different pros and cons. However, top-3 approaches significantly outperformed other teams, with 0.03-0.16 observed in OvAcc (Wilcoxon signed-rank test, $p<0.05$). Interestingly, only these top-3 methods implemented weighted general union loss, indicating the importance of adopting specific optimization functions. Furthermore, the detection rate of the first and second teams was significantly higher than that of the third team (Wilcoxon signed-rank test, $p < 0.05$). The use of optimization functions with an emphasis on the connectivity of airway branches has significantly improved the results of airway segmentation. This enables the AI model to better reflect the anatomical architecture of airway trees, leading to more potential clinical outcomes.

Regarding the capacity of airway modelling, all models achieved better performance (OvAcc) on COVID-19 cases than fibrosis cases (Wilcoxon signed-rank test, $p<0.05$). In addition, there exists a significant difference (Wilcoxon signed-rank test, $p<0.05$) in the DLR and DBR across the COVID-19 and fibrosis datasets. This disparity between the two datasets may arise from the distinct pathological manifestations of COVID-19 and fibrotic lung disease, indicating the HRCT collected from fibrosis patients might inherently contain more complex structures for AI applications. Conversely, the presence of airway leakages and missing segments might be more prominent in the COVID-19 dataset, which is 1~3% higher than that of fibrosis patients. This aligns with the higher DBR and DLR observed in the COVID-19 dataset; as more branches were detected, the model became error-prone.

**Overall performance of mortality prediction.**

Task II is structured as a binary classification to predict patient mortality status after 63 weeks. The validation set was maintained balanced, consisting of 26 samples for label 0 (alive) and 26 samples for label 1 (deceased). Interestingly, the test dataset was randomly selected and 87% of the cases were survivors who surpassed the 63-week threshold. This imbalance was highlighted by team DJ_92, who achieved first place in the validation stage while the fourth in the test set.

The results in Table. 4 indicated a sustained preference for radiomics features in extracting quantitative imaging biomarkers, attributed to their interpretability and limited GPU memory demands. Nonetheless, a degree of inconsistency in their performance is observable, as evidenced by the diverse results (Table. 5). This inconsistency can be traced back to the reliance on ROI-based radiomics features by the majority of Task II participants. Since these features are heavily dependent on precise segmentation, varying region selection approaches can lead to diverse results. Moreover, CNN-based methods come with inherent limitations. Given the substantial size of whole volumetric CT scans, it becomes nearly impractical to directly input the original 3D CT image into a deep learning network. To address this challenge, down-sampling (by Tastefish team) and cropping (by junqiangmler) were implemented to reduce the GPU memory cost.

ROC curves of the top-5 submissions were plotted in Fig. 4 to show the diagnostic ability of the capacity of the binary classifier system when its discrimination threshold is varied. This was incorporated to assess the impact of different thresholds on the outcome. The uAI-Team's model exhibits the best performance (0.8 AUC scores) among the submitted models in terms of the ability to discriminate the patient mortality status after 63 weeks. However, it is noteworthy that the uAITeam requires information on lesion characteristics.

**Delving into mispredictions in airway segmentation.**

Figure 5 presents a visual comparison of the performance of eight top-performing teams in airway modelling on the fibrosis test set (N=90), using heatmaps to illustrate the differences between the actual ground truth and the predicted outputs ($\mathcal{H} = (Y - X)$). The accumulation maps of the prediction and ground truth were first generated by adding different slices along the vertical axis, followed by calculating the residual map between these accumulated maps. These heatmaps are colour-coded: warm colours indicate areas where false negatives occurred (i.e., the model failed to predict an airway where one existed), and cold colours represent false positives (i.e., the model predicted an airway where none existed).

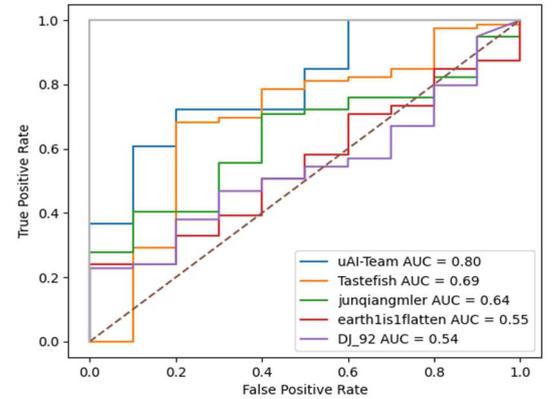

The visual analysis highlights specific trends in the predictive capabilities of different teams. For instance, the majority of the mispredictions by the **MedibotTeam** and **IMR** were false positives, which were predominantly found in the smaller-sized branches. This tendency contributed to their relatively high airway leakage rates, with MedibotTeam at 0.08 and IMR at 0.11, which is higher compared to the other teams. Furthermore, the heatmaps state the IMR team's scoring patterns, explaining why they achieved high scores for DBR and DLR metrics, while their IoU scores were comparatively lower. The white bounding boxes in the heatmap of the IMR team reveal that, despite the team's proficiency in identifying terminal and small branches, their method was less accurate when it came to medium-sized branches and the trachea. This insight into the distribution of prediction errors helps us understand the strengths and weaknesses of the different models in predicting airway structures. For team **Infervision**, the heatmap exhibits fewer cold regions within the lung lobes, indicative of fewer false negatives, which is consistent with their significantly (Wilcoxon signed-rank test, $p<0.05$) low airway leakage rate of 0.601 compared with MedibotTeam and IMR. The heatmap of **Sanmed AI** demonstrates a more mixed pattern of warm and cold colours. The deep colour around the lobar bronchus (labelled by the black bounding box) indicates comparable airway leakages, while that of the trachea (white bounding box) refers to heavy neglect of the main trachea. Interestingly, the false positive region (black arrow at the lower left corner) presents the misprediction of artefacts of scanning devices, indicating the lack of keeping the largest connective component postprocessing algorithm. Conversely, the result of team **Gexing** reflects false positive predictions of the trachea and medium-sized branches, while their model ignores certain small and terminal branches. Team **DJ_92** and **Riipl** present similar patterns with fewer false positive predictions, which corresponds to their high precision scores and low branch scores (DLR and DBR). For the models towards the lower end of the performance, e.g., **the Earth1is1flatten,** their heatmaps were dominated by false negative predictions and some false positive predictions were blocked. The heatmaps offer a visual narrative that complements the numeric data in Table 3 and S-Table 2, providing a deeper understanding of the strengths and weaknesses of different models.

Fig. 4. ROC curves of the top-5 teams in Task II.

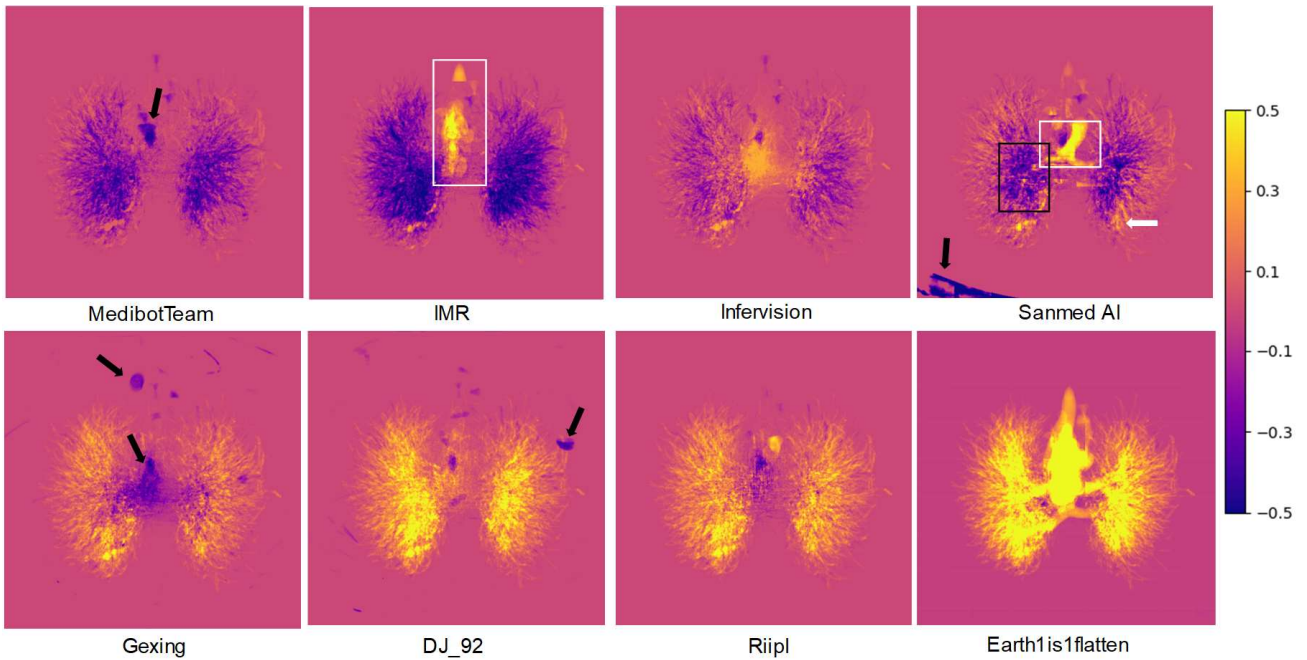

Fig. 5. Heatmap of the residual between the ground truth and the prediction on the accumulated axis plane. The warm colours indicate false negative predictions (missing prediction), and the cold colours refer to false positive predictions (leakages).

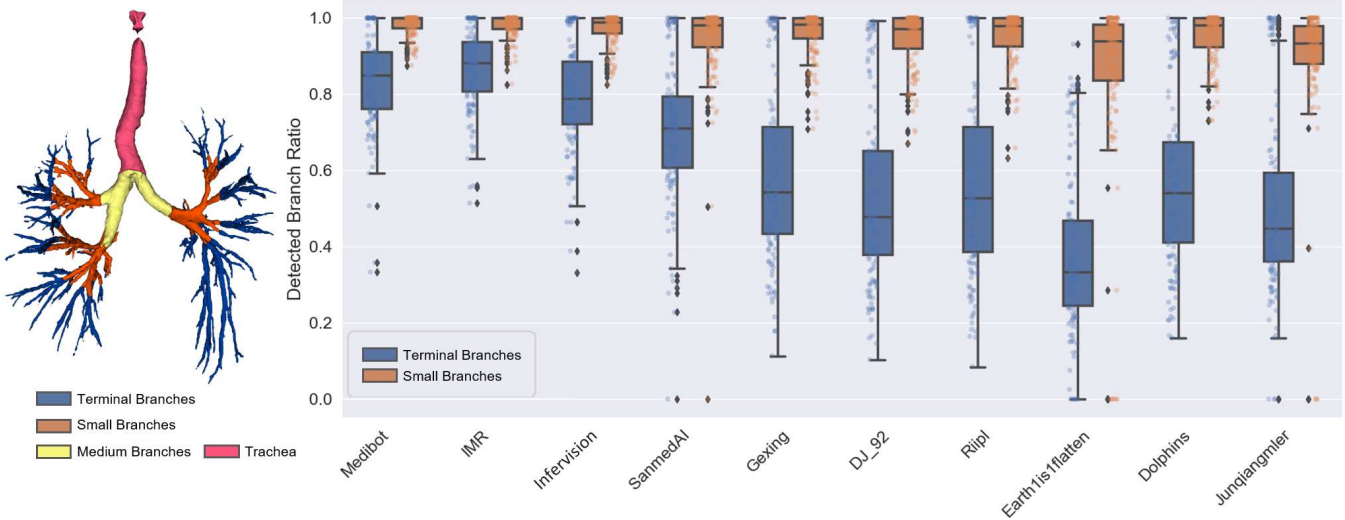

Fig. 6. Boxplot of the detected branch rate for terminal and small branches.

**Rethinking the airway modelling, how do models perform on different-sized branches?**

Airway trees can be split into different groups based on their anatomical attributes, e.g., their size. To investigate the detailed performance of these models, airway trees were first grouped into terminal (TB), small (SB), medium (MB), and large branches, followed by the assessment of detected branch rate according to (Nan et al., 2023) (details in S-Table 5). Specifically, terminal branches have a radius ranging from 0-2mm, small branches have a radius of 2-4mm, medium branches are 4-8mm in radius, and large branches have a radius greater than 8mm.

As S-Table 5 shows, most approaches achieved better performance (detection rate > 0.95) on identifying medium and large branches than small and terminal branches (Wilcoxon signed-rank test, $p<0.05$). The detection rate for small branches was comparable (ranging from 0.86 to 0.98), while that for terminal branches was various (Fig. 6). For instance, IMR achieved the highest detection rate of 0.8678 for terminal branches, while the lowest figure among the top-10 teams was only 0.3822. This indicates that most deep learning-based methods could well segment the majority of airway trees in lung fibrosis patients, while the recognition of terminal branches remains to be further improved. Additionally, although IMR has a high detection rate for the trachea, it inevitably missed part of the main bronchus, which is the reason why their model simultaneously achieved high detection rate (S-Table 5) and high false negative predictions (Figure 5) of the trachea.

## How do models perform on unharmonised data?

In clinical practice, HRCT scans were collected by different hospitals with various acquisition protocols. This includes different acquisition devices, different reconstruction kernels, etc (Nan et al., 2022). Here, we mainly investigate how models perform on unharmonised data by testing on the manual flipped HRCT scans. The original test fibrosis scans (n=90) were flipped through three different axes along z, y, and x. Subsequently, we adopted top-5 approaches in AIIB23 to assess how these models perform on the unharmonised data (in Table 6).

Table. 6. Model performance on unharmonised data.

|  | Data Process | IoU | DLR | DBR | ALR | AMR |
|---|---|---|---|---|---|---|
| MedibotTeam | X-Flip | 0.8971 | 0.9109* | 0.8700* | 0.0779* | 0.0339 |
|  | Y-Flip | 0.8965 | 0.9066* | 0.8635* | 0.0743* | 0.0378 |
|  | Z-Flip | 0.8970 | 0.8964* | 0.8509* | 0.0713* | 0.0398 |
| IMR | X-Flip | 0.8698 | 0.9255 | 0.8952 | 0.1063 | 0.0389 |
|  | Y-Flip | 0.8665 | 0.9283 | 0.9003 | 0.1074 | 0.0418 |
|  | Z-Flip | 0.8717 | 0.9079* | 0.8717*` | 0.0965* | 0.0454* |
| Infervision | X-Flip | 0.8944 | 0.8855 | 0.8352 | 0.0605 | 0.0522 |
|  | Y-Flip | 0.8906 | 0.8830 | 0.8322 | 0.0571 | 0.0591 |
|  | Z-Flip | 0.8919 | 0.8793 | 0.8289 | 0.0562 | 0.0584 |
| Sanmed_AI | X-Flip | 0.8828 | 0.8418 | 0.7833 | 0.0742 | 0.0526 |
|  | Y-Flip | 0.3529* | 0.4931* | 0.4632* | 0.2260* | 0.6275* |
|  | Z-Flip | 0.5196* | 0.5893* | 0.5366* | 0.0532* | 0.4571* |
| Gexing | X-Flip | 0.8994 | 0.7897* | 0.6871* | 0.0426 | 0.0625 |
|  | Y-Flip | 0.8871* | 0.7918* | 0.6923* | 0.0383* | 0.0791* |
|  | Z-Flip | 0.8887* | 0.7851* | 0.6853* | 0.0383* | 0.7750* |

* indicates significant changes (Wilcoxon signed-rank test, p<0.05) compared with the performance on original test cases, the green colour indicates positive trends while the red colour refers to negative trends.

Among the top five approaches, the model developed by team Infervision exhibited the most robust performance, with its predictions on unharmonized data showing no significant differences from those on the original test set. In addition, the IoU scores of all top three methods remained steady, while those of team Sanmed_AI and Gexing were fluctuant. Unexpectedly, the DLR and DBR scores for MedibotTeam experienced a notable decline, in contrast to team IMR, whose scores only decreased significantly on Z-Flipped data. Interestingly, most models (MedibotTeam, IMR, Sanmed AI and Gexing) presented a significant reduction in leakages on unharmonised data, suggesting a potential advantage of test-time augmentation. In contrast to other teams that experienced a slight decline in DLR and DBR scores, ranging from 1 to 3%, the performance of Sanmed AI on Y-Flip and Z-Flip deteriorated significantly. Specifically, Sanmed AI encountered steep falls in IoU (30-50%), DBR and DLR (nearly 40%) scores on the Y and Z flipped datasets. The robust performance of team Infervision on unharmonised data suggested that applying appropriate data augmentation techniques could effectively mitigate biases related to the data orientation.

## How do models perform on low-quality data?

Low-quality data, characterized by noise, inconsistencies, missing values, or bias, can significantly hinder a model's ability to generate accurate and reliable predictions. When fed with such imperfect datasets, models can overfit the noise, capturing irrelevant patterns that don't generalize well to real-world scenarios. Here we test the capabilities of top-5 models when facing low-quality data by introducing manual noises and down-sampling (Table. 7). For noise images, we randomly introduced Gaussian noises to the original HRCT scans, and we randomly down-sampled (reduction ratio $\in [0.5, 1]$) along the z-axis (reduce the number of slices) to obtain the inconsistent data.

Table. 7. Model performance on low-quality data

|  | Data Transfer | IoU | DLR | DBR | ALR | AMR |
|---|---|---|---|---|---|---|
| MedibotTeam | Down sample | 0.8889 | 0.8805* | 0.8421* | 0.0824 | 0.0391* |
|  | Noise | 0.8952 | 0.8552* | 0.7885* | 0.0636* | 0.0484* |
| IMR | Down sample | 0.8657 | 0.8955* | 0.8657* | 0.1057 | 0.0440* |
|  | Noise | 0.8710 | 0.8787* | 0.8252* | 0.0848* | 0.0558* |
| Infervision | Down sample | 0.8856* | 0.8407* | 0.7943* | 0.0583 | 0.0633* |
| | Noise | 0.8941 | 0.8457* | 0.7746* | 0.0476* | 0.0637* |
| Sanmed_AI | Down sample | 0.8709 | 0.8008* | 0.7482* | 0.0825 | 0.0687* |
| | Noise | 0.8815 | 0.7578* | 0.6662* | 0.0536* | 0.0719* |
| Gexing | Down sample | 0.8945 | 0.7548* | 0.6662* | 0.0413 | 0.0687* |
| | Noise | 0.8821* | 0.6949* | 0.5791* | 0.0373* | 0.0851* |

* indicates significant changes (Wilcoxon signed-rank test, p<0.05) compared with the performance on original test cases, the green colour indicates positive trends while the red colour refers to negative trends.

In particular, there was a general reduction of branch metrics and airway missing rates in all five approaches (p<0.05). Interestingly, the airway leakage ratio of all models was significantly decreased on the noise dataset, which might be associated with the increase of false negative predictions (more terminal branches were missed in predictions). The prevalent reduction also suggests that there is an urgent need to develop a robust model that can well address the low-quality data.

### Beyond segmentation, what is more?

To further investigate the clinical value of airway-derived biomarkers, we conducted survival analyses for patients with fibrotic lung diseases. Data used in this analysis were collected from AIPFR, with 203 HRCT scans available for airway modelling. In this section we mainly focused on a primary airway-derived biomarker, the total airway volume (TAV), which is the total volume counts of the entire airway trees against the total volume of lung lobes. In addition to the proposed TAV, probabilities of the prediction from Task-II teams (top three teams including uAI, Tastefish, and Junqiangmler) were involved in survival analysis (denoted as Prob_II1, Prob_II2, Prob_II3).

The main steps for survival analysis can be summarized as: The computational biomarkers (TAV, Prob_II1, Prob_II2, and Prob_II3) were first analysed to find out if they were associated with the mortality. Then, these biomarkers would be analysed with the existing biomarkers. This includes the imaging biomarkers SOFIA (Walsh et al., 2018), average fibrosis score (given by three senior radiologists to describe the total fibrosis extent), and physiological measurements such as FVC percentage (FVC%), to explore its independence when predicting mortality. It is of note that all these survival analyses are performed through the Cox Proportional-Hazards model via StataMP V17.0.

The univariate results are shown in Table. 8 suggest that all computational biomarkers (TAV, Prob_II1, Prob_II2, and Prob_II3) were significantly associated with the mortality (p<0.05). For bivariate analysis with CPI, total fibrosis extent, FVC% or SOFIA, only TAV and Prob_II1 were independent variables (p<0.05) that were predictive of mortality. The relationship between TAV, Prob_II1, Prob_II2, Prob_II3, and SOFIA was reported through multivariate analysis. Among these five AI-based computational biomarkers, TAV, Prob_II1 and SOFIA were independent variables in predicting mortality (p<0.05), while Prob_II2 and Prob_II3 did not demonstrate independent predictive value for mortality at the statistical threshold (p≥0.05). Based on this, further multivariate survival analyses were conducted among TAV, ProbII1, SOFIA, AvgFib and FVC%. The results showed that the TAV has demonstrated a highly significant association with mortality, evidenced by a p-value of less than 0.0001 and a hazard ratio (HR) of 1.83. Conversely, Prob_II1, while indicative of some risk with an HR of 1.70, does not reach statistical significance due to a p-value of 0.089. SOFIA trended toward statistical significance, suggesting a marginal effect on mortality that does not achieve conventional statistical confirmation. The total fibrosis extent was observed with no significant independent predictive power for mortality. In contrast, FVC%, with a p-value of 0.025 and an HR of 0.35 (95% CI: 0.14-0.88), is statistically significant and suggests a substantial reduction in mortality risk associated with each unit increase.

Table. 8. Survival analysis for image-derived biomarkers

| Methods | Variables | P Value | HR (95% CI) | Variables | P Value | HR (95% CI) |
|---|---|---|---|---|---|---|
| Univariate (N=203) | TAV | <0.0001 | 2.58 (2.05-3.24) | Prob_II2 | 0.003 | 1.80 (1.22-2.66) |
| | Prob_II1 | <0.0001 | 3.06 (1.86-5.05) | Prob_II3 | 0.008 | 2.04 (1.21-3.43) |
| Bivariate with AvgFib (N=200) | Prob_II1 | 0.037 | 1.83 (1.04-3.24) | Prob_II3 | 0.025 | 1.84 (1.08-3.15) |
| | AvgFib | <0.0001 | 1.04 (1.02-1.05) | AvgFib | <0.0001 | 1.04 (1.03-1.06) |
| | Prob_II2 | 0.156 | 1.34 (0.89-2.00) | TAV | <0.0001 | 2.10 (1.57-2.82) |
| | AvgFib | <0.0001 | 1.04 (1.03-1.06) | AvgFib | 0.001 | 1.03 (1.01-1.04) |
| Bivariate with FVC% | Prob_II1 | <0.0001 | 2.52 (1.50-4.24) | Prob_II3 | 0.195 | 1.43 (0.83-2.47) |

|  | (N=203) | FVC% | <0.0001 | 0.19 (0.08-0.45) | FVC% | <0.0001 | 0.17 (0.07-0.43) |
|---|---|---|---|---|---|---|---|
|  |  | Prob_II2 | 0.029 | 1.55 (1.05-2.31) | TAV | <0.0001 | 2.35 (1.80-3.07) |
|  |  | FVC% | <0.0001 | 0.18 (0.08-0.42) | FVC% | 0.004 | 0.27 (0.11-0.66) |
| Bivariate with SOFIA (N=203) |  | Prob_II1 | <0.0001 | 2.59 (1.52-4.41) | Prob_II3 | 0.057 | 1.67 (0.98-2.85) |
|  |  | SOFIA | <0.0001 | 1.36 (1.20-1.55) | SOFIA | <0.0001 | 1.38 (1.22-1.57) |
|  |  | Prob_II2 | 0.104 | 1.40 (0.93-2.10) | TAV | <0.0001 | 2.19 (1.67-2.89) |
|  |  | SOFIA | <0.0001 | 1.37 (1.20-1.56) | SOFIA | 0.001 | 1.26 (1.10-1.44) |

| Methods | Variables | P Value | HR (95% CI) |
|---|---|---|---|
| Multivariate (n=203) | TAV | <0.0001 | 2.10 (1.57-2.81) |
|  | Prob_II1 | 0.007 | 2.22 (1.24-3.99) |
|  | Prob_II2 | 0.735 | 0.924 (0.59-1.46) |
|  | Prob_II3 | 0.471 | 1.24 (0.69-2.26) |
|  | SOFIA | 0.005 | 1.23 (1.07-1.42) |
| Multivariate (n=200) | TAV | <0.0001 | 1.83 (1.35-2.49) |
|  | Prob_II1 | 0.089 | 1.70 (0.92-3.13) |
|  | SOFIA | 0.052 | 1.17 (1.00-1.38) |
|  | AvgFib | 0.163 | 1.01 (0.99-1.03) |
|  | FVC% | 0.025 | 0.35 (0.14-0.88) |

**Reproducibility of radiomic features.**

Radiomic features play a crucial role in explainable clinical diagnosis and prognosis tasks, which refer to a method for the quantitative description of medical images. The radiomic features were believed to be able to capture the unknown patterns and characteristics that were not observed by clinicians. However, existing studies have found that these radiomic features had low reproducibility and the reliability of these features on lung fibrosis disease was not clear. Although the overall accuracy (OvAcc) of the top-2 teams in Task I present statistical similarity (p>0.05), it is unclear whether the extracted radiomic features remained the same. This section first analysed the variance of the radiomic features between different AI models and then reported the applicable radiomic features for clinical prognostication.

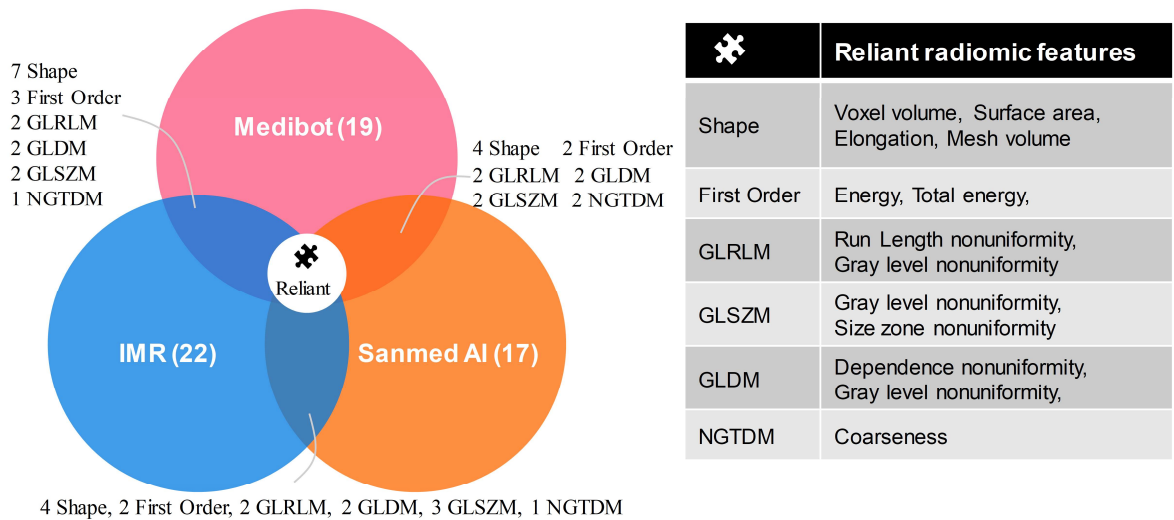

Fig. 7. Reliant radiomic features in survival status prediction.

We first analysed the significance between the radiomic features extracted from MedibotTeam, IMR and Sanmed AI (the 1st, 2nd and 4th team in Task I, respectively), followed by implementing Cox Hazard analysis to find variables related to the mortality. An overall of 107 radiomic features were extracted for each case, including 14 shape features, 18 first-order features, 24 Grey Level Co-occurrence Matrix features (GLCM), 14 Grey Level Dependence Matrix (GLDM), 16 Grey Level Run Length Matrix (GLRLM), 16 Grey Level Size Zone Matrix (GLSZM), and 5 Neighbouring Grey Tone Difference Matrix Features (NGTDM). Interestingly, most of these features (>86) showed significant differences (Wilcoxon signed-rank test, p<0.05) among different models, even though there were no statistical differences (p<0.05) in the overall accuracy between MedibotTeam and IMR. Among 107 features, 19, 22, and 17 features were observed to be associated with mortality (p<0.05 in the univariate Cox proportional hazards Model), while there were 13 common features for these three models (Fig. 7).

**Limitations and future research directions.**

Although efforts have been made, several issues remain to be addressed in future studies. (1) Deficient central line extraction algorithm. Due to the complex structure of airway trees, airway central lines estimated by skeleton algorithms are error-prone, which leads to the miscalculation of branch metrics (e.g., the number of branches estimated by algorithms was greater than its real value). (2) Insufficient robustness. As Tables 6 and 7 demonstrated, most approaches suffered from low generalization ability when facing noisy and low-resolution CT scans. This indicates the urgent need for developing a generalizable model for airway extraction, addressing the unharmonised data and low-quality data. (3) Lack of in-depth clinical analysis. Most approaches ignored the clinical application of airway modelling, obstructing the utilization of AI in clinical practice. This study has revealed the perspective of airway-derived imaging biomarkers in disease prognostication whereas further in-depth analyses could be conducted. For instance, investigating the effects of different branches, airway lumen and wall measurements.

## Conclusion

This paper presented details of the AIIB23 challenge with the state-of-the-art approaches for airway modelling and mortality prediction in patients with fibrotic lung disease. The results showed the domination of the CNN-based method in medical image segmentation and surprisingly, highlighted the importance of using explainable image-derived features for mortality prediction. The highest-rank team in Task I achieved an overall score of 0.9185 and an average inference speed of 43.8s per scan, and that of Task II achieved 0.8493 F1-score in mortality prediction. We found that (1) the capacity of models (for airway segmentation) could be enhanced by introducing the weighted general union loss and continuity loss; (2) the decreased effectiveness of AI models when facing low-quality data underscores the urgent need to develop models empowered with robust generalization capabilities; (3) there were 13 reliant radiomic features related to the mortality and could be used for prognostication tasks; (4) the total airway volume TAV, is a strong imaging-biomarker for prognosis of patients with fibrosis compared with existing AI-based biomarkers and clinical measurements. The study also revealed potential research directions such as the centreline extraction algorithm could be improved to reduce evaluation errors.

## Acknowledgement

This study was supported in part by the ERC IMI (101005122), the H2020 (952172), the MRC (MC/PC/21013), the Royal Society (IEC\NSFC\211235), the NVIDIA Academic Hardware Grant Program, the SABER project supported by Boehringer Ingelheim Ltd, NIHR Imperial Biomedical Research Centre (RDA01), Wellcome Leap Dynamic Resilience, and the UKRI Future Leaders Fellowship (MR/V023799/1). Send correspondence to G. Yang at g.yang@imperial.ac.uk, Y. Nan at y.nan20@imperial.ac.uk, and X. Xing at x.xing@imperial.ac.uk